\documentclass[11pt,b5paper,twoside]{article}

\usepackage[cp1251]{inputenc}
\usepackage[T2A]{fontenc}
\usepackage[russian]{babel}

\usepackage[dvips]{graphicx}
\usepackage{amsmath}
\usepackage{amssymb}
\usepackage{amsxtra}
\usepackage{amsthm}
\usepackage{latexsym}
\usepackage{array}
\usepackage{multirow}
\usepackage{hhline}

\newtheorem{remark}{Замечание}
\newtheorem{proposition}{Предложение}

\newcommand {\mS}{\mathcal{S}}
\newcommand {\bs}{\boldsymbol}
\newcommand{\Res}{\mathop{\rm Res }\nolimits}

\begin{document}

\begin{center}

\large{{\bf УСЛОВИЯ СУЩЕСТВОВАНИЯ\\ПЕРИОДИЧЕСКИХ ДВИЖЕНИЙ\\ГИРОСТАТА КОВАЛЕВСКОЙ В ДВОЙНОМ ПОЛЕ\footnote{Работа выполнена при финансовой поддержке гранта РФФИ и
Администрации Волгоградской области № 10-01-97001.}}}

\vspace{5mm}

\normalsize

{\bf И.И. Харламова, Г.Е. Смирнов}

\vspace{4mm}

\small

Волгоградская академия государственной службы

Россия, 400131, Волгоград, ул. Гагарина, 8

E-mail: irinah@vags.ru

Московский государственный университет им. М.В. Ломоносова

Москва, Воробьевы горы

E-mail: glebevgen@yandex.ru

\end{center}

\begin{flushright}
{\it Получено 13 августа 2010 г.}
\end{flushright}

\vspace{3mm}

\footnotesize{Изучаются особые периодические движения (критические точки ранга 1
интегрального отображения), найденные в работе М.П.\,Харламова (Механика твердого тела,
вып.~37, 2007) в интегрируемой задаче о движении
гиростата в двойном поле при условиях Ковалевской на моменты
инерции. Исследованы возможные перестройки внутри
множества этих решений в зависимости от существенных параметров --
одного интегрального и двух физических. Получены аналитические
уравнения разделяющего множества и его особенностей, указано
количество возникающих областей с различным набором решений. Найден
образ разделяющего множества в пространстве параметров, задающих
бифуркационные диаграммы на изоэнергетических уровнях. Вычисления,
связанные с преобразованиями многочленов высоких степеней, выполнены
в компьютерной системе Mathematica 7.
}

%\tableofcontents

\section{Введение}\label{sec1}
В невырожденной интегрируемой
гамильтоновой системе критические точки ранга~1 (так называемые
особые периодические движения) организованы в подсистемы с одной
степенью свободы. Поэтому в принципе соответствующие траектории
образуют однопараметрические семейства (при фиксированных физических
параметрах задачи). Сам факт построения периодических решений всегда имеет в динамике
особое значение. Но важность исследования таких семейств, кроме этого,
состоит также и в том, что в системах с
тремя степенями свободы их бифуркации порождают перестройки типов
плоских сечений бифуркационных диаграмм задачи в целом. В случае
Ковалевской--Реймана--Семенова-Тян-Шанского \cite{ReySem} при
отсутствиии гиростатического момента $\lambda$ (волчок в двойном
поле) явное представление зависимости особых периодических движений
(ОПД) от одного интегрального параметра позволили
получить \cite{Kh362} полную классификацию бифуркационных диаграмм
$\Sigma_h$ интегрального отображения на изоэнергетических
поверхностях $H=h$, т.е. построить разделяющее множество на
плоскости $(h,\gamma)$, где $\gamma$ -- единственный существенный
физический параметр задачи (отношение напряженностей силовых полей).
При $\lambda \ne 0$ (гиростат в двойном поле) все критические точки
ранга 1 также найдены \cite{Kh37} и уравнения соответствующих ОПД
интегрируются в эллиптических функциях. Однако эти решения удалось
выписать лишь в зависимости от двух нефизических параметров,
связанных уравнением высокой степени. Через эти же параметры
выражены и постоянные общих интегралов. Это удобно для численного
построения плоских диаграмм $\Sigma_h$, так как при заданном $h$
эффективно вычисляются все граничные точки одномерных сегментов.
Попытки явно описать соответствующие разделяющие поверхности в
пространстве $(h,\gamma,\lambda)$, названные атласом бифуркационных
диаграмм \cite{Ryab37}, еще не привели к окончательному результату.
В настоящей работе выполнено исследование условий существования и
бифуркаций ОПД, выписаны явные уравнения соответствующих разделяющих
поверхностей. Отметим, что многие вычисления выполнены в
компьютерной системе Mathematica 7 (Academic License~\#~L3298-7174)
ввиду крайне высоких вычислительных сложностей. При этом во
всех случаях удалось провести исследование полностью, доказать все
утверждения о количестве и характере решений тех или иных уравнений.
Несмотря на использование компьютерных технологий, все рассуждения,
представленные ниже, являются доказательствами, поскольку описаны
все этапы вычислений, предъявлены промежуточные результаты, и любое
вычисление при необходимости может быть легко повторено. Ряд
результатов вывести <<вручную>> уже нельзя. В то же время, в
большинстве случаев получить ответ напрямую не помогает и компьютер
-- необходимо увидеть и задать в компьютерной системе нужную
подстановку. Поэтому, несмотря на фантастические возможности систем
аналитических вычислений, роль исследователя по-прежнему остается
решающей.

\section{Исходные уравнения}\label{sec2} Пусть $\bs \alpha$, $\bs
\beta$ -- характеристические векторы взаимно ортогональных силовых
полей с модулями ${|\bs \alpha|=a}$, ${|\bs\beta|=b}$ и
\begin{equation}\label{eq2_1}
    a>b>0.
\end{equation}
Пусть $\bs \omega$ -- угловая скорость тела, $\lambda$ --
единственная отличная от нуля, осевая компонента вектора
гиростатического момента. Выражения фазовых переменных через
вспомогательную переменную $w=\omega_1^2+\omega_2^2 \geqslant 0$
описываются уравнениями (30) работы \cite{Kh37}. В этих выражениях
фигурируют физические параметры $\lambda$, $p$,
$r$ ($p=\sqrt{a^2+b^2}>r=\sqrt{a^2-b^2}>0$), а также два параметра $\sigma, u$, первый из
которых есть неопределенный множитель Лагранжа в линейной
зависимости дифференциалов общих интегралов, и потому сам является
частным интегралом на рассматриваемых траекториях, а второй --
вспомогательный параметр, введенный в процессе вычислений с целью
получения простых формул и связанный с $\sigma$ уравнением
\begin{equation}\label{eq2_2}
\begin{array}{l}
  L=\lambda^2(\lambda^2+\sigma)^2 u^5+(\lambda^2+\sigma)
[2p^2\lambda^4-(\lambda^2+\sigma)^3 \sigma]\sigma u^4+ \\
  \qquad +r^4\lambda^6\sigma^2 u^3+ 2 r^4 \lambda^4
\sigma^4(\lambda^2+\sigma)^2u^2-r^8\lambda^8 \sigma^6=0.
\end{array}
\end{equation}
Зависимость $w$ от времени описывается уравнением
\begin{equation} \label{eq2_3}
\displaystyle{\big(\frac{dw}{dt}
\big)^2=-\frac{\lambda^2}{4\sigma^2} P_4(w) ,}
\end{equation}
где $P_4(w)=P_+ (w) P_- (w)$ и
\begin{equation}\notag
\begin{array}{c}
    \displaystyle{P_{\pm}(w) = w^2+2\sigma^2\frac{u \pm
r^2\lambda^2}{\lambda^2 u}w +
\frac{\sigma[u^3-(\lambda^2+\sigma)\sigma^2u^2+r^4\lambda^4
\sigma^3]}{(\lambda^2+\sigma)\lambda^2u^2}.}
\end{array}
\end{equation}
В частности, условия вещественности решений имеют вид системы
неравенств
\begin{equation}\label{eq2_4}
    P_4(w) \leqslant 0, \qquad w \geqslant 0,
\end{equation}
исследование совместности которых будем проводить в плоскости
$(\sigma, \lambda)$ с учетом зависимости от $a,b$. Хотя существенным
является лишь отношение $\gamma=b/a \in (0,1)$, сохраним в общих
уравнениях оба параметра для возможности в будущем использовать
любые предельные переходы. Отметим, что во всех приведенных формулах
встречаются лишь четные степени $\lambda$, поэтому достаточно
рассмотреть полуплоскость $\lambda \geqslant 0$. Ниже все
утверждения формулируются для этой полуплоскости, что особо уже не
оговаривается. При фиксированных $\sigma, \lambda$ имеется до пяти
значений $u$, удовлетворяющих \eqref{eq2_2}. Некоторым из них
соответствуют решения уравнения \eqref{eq2_3} при условиях
\eqref{eq2_4}, причем таких решений может быть и несколько, если
множество \eqref{eq2_4} несвязно. Пусть
$\mathcal{P}(\sigma,\lambda,u)$ -- набор решений \eqref{eq2_3} как совокупность
геометрических объектов на плоскости $(w,\dot w)$ и пусть
${\mathcal{P}(\sigma,\lambda)=\cup_u
\,\mathcal{P}(\sigma,\lambda,u)}$. Применительно к задаче
количественной классификации ОПД назовем {\it разделяющим множеством}
подмножество $\mS$ в плоскости $(\sigma,\lambda)$, при переходе
через которое меняется набор $\mathcal{P}(\sigma,\lambda)$. Сразу же
отметим часть $\mS_0$ разделяющего множества $\mS$, состоящую из
координатных осей и особой параболы $\Pi: \, \lambda^2+\sigma=0$. На
этом множестве выражения для явных решений, выписанные в \cite{Kh37}, имеют особенности. Как показано
в \cite{KhHMJ}, особенность на параболе устранима предельным переходом,
однако перестройки в множестве $\mathcal{P}(\sigma,\lambda)$ при
этом не исключены.

\section{Первое разделяющее множество}\label{sec3} Периодические
решения претерпевают бифуркации при переходе через неподвижные
точки, которые соответствуют кратному корню многочлена $P_4$.
Дискриминанты многочленов $P_{\pm}$ имеют вид
\begin{equation}\notag
\begin{array}{l}
\displaystyle{\frac{\sigma}{\lambda^4(\lambda^2+\sigma)u^2}D_{\pm},}\qquad
D_{\pm} = [r^2\lambda^2\sigma \pm
(\lambda^2+\sigma)u]^2\sigma^2-\lambda^2 u^3.
\end{array}
\end{equation}

\begin{remark}[1] {\it Здесь и далее в результантах и дискриминантах нас интересует лишь их
зависимость от параметров $($возможность обращения в нуль$)$ и, в
отдельных случаях, их знак. Поэтому, для простоты записи, пишем
выражения, имея в виду равенство с точностью до положительного
числового множителя.}
\end{remark}

В соответствии с этой договоренностью, результанты многочленов
$D_{\pm}$ с многочленом $L$ по переменной $u$ равны $r^{16} (p^2 \pm
r^2)^3\lambda^{30} \sigma^{19} (\lambda^2+\sigma)^3$. Эти значения
не обращаются в нуль за пределами $\mS_0$. Следовательно, кратный
корень $P_4$ может быть лишь общим корнем $P_+$ и $P_-$. Тогда, как
и должно быть в неподвижной точке, этот корень нулевой. Поэтому
\begin{equation} \label{eq3_1}
P_0(u)=u^3-(\lambda^2+\sigma)\sigma^2u^2+r^4\lambda^4 \sigma^3=0.
\end{equation}
Приравниваем к нулю результант многочленов $L$ и $P_0$ по $u$, получим
\begin{equation}\notag
\begin{array}{l}
r^{16}\lambda^{24}\sigma^{15}(\lambda^2+\sigma)^3
\left\{ r^8 + r^4[6 p ^2 \lambda^4 + \lambda^8 - 4\lambda^2(p^2 +
\lambda^4) \sigma - 2(2 p^2
+ \lambda^4)\sigma^2 +\right.\\[3mm]
\left.  + 4\lambda^2 \sigma^3 + 2 \sigma^4] -
[\sigma(\lambda^2 + \sigma)-2 p^2]^2 [2 p^2 \lambda^4 - \sigma^2
(\lambda^2 + \sigma)^2)] \right\}=0.
\end{array}
\end{equation}
Вне $\mS_0$ достаточно найти нули выражения в фигурной скобке,
которое разлагается на множители в виде $Q_1 Q_2 Q_3 Q_4$. Поэтому
первое разделяющее множество (ПРМ) в составе $\mS$ -- четыре кривые
\begin{equation}\notag
\begin{array}{l}
Q_1=(a-b)^2+(a+b)\lambda^2-(\lambda^2+\sigma)\sigma=0,\\
Q_2=(a+b)^2+(a-b)\lambda^2-(\lambda^2+\sigma)\sigma=0,\\
Q_3=(a+b)^2-(a-b)\lambda^2-(\lambda^2+\sigma)\sigma=0,\\
Q_4=(a-b)^2-(a+b)\lambda^2-(\lambda^2+\sigma)\sigma=0.
\end{array}
\end{equation}
Сами эти кривые будем обозначать так же, как и левые части уравнений
-- через $Q_1 - Q_4$, а их объединение -- через $\mS_1$. Значения
энергии в неподвижных точках тела в двойном поле хорошо известны
(см., например, \cite{Kh34}). В работе \cite{Kh37} выполнен сдвиг
энергии на константу. Примем здесь стандартное выражение интеграла
энергии
\begin{equation}\notag
    H=\omega_1^2+\omega_2^2+\frac{1}{2}\omega_3^2-(\alpha_1+\beta_2),
\end{equation}
тогда его критические значения таковы
\begin{equation}\label{eq3_2}
    h= \pm a \pm b.
\end{equation}
Покажем, как получить их в точках ПРМ. Исключая старшие степени $u$
в многочленах $P_0$ и $L$, на кривых $Q_i$ найдем
\begin{equation}\label{eq3_3}
u_{1,2}=(a \pm b)\lambda^2\sigma, \qquad u_{3,4}=(-a \pm
b)\lambda^2\sigma.
\end{equation}
Подставим в функцию $H$ значения фазовых переменных, полученные из выражений (8), (30) работы \cite{Kh37}:
\begin{equation}\label{eq3_4}
h=\displaystyle{-\frac{u^3 \lambda^2 -
  r^4 \lambda^4 \sigma^3 (\lambda^2 + 2 \sigma) +
  u^2 \sigma^2 (\lambda^2 + \sigma) (\lambda^2 +
     2 \sigma)}{2 u^2 \lambda^2 \sigma (\lambda^2 + \sigma)}}.
\end{equation}
Исключим отсюда $u^3$ с помощью условия \eqref{eq3_1}, и
подставим $u_i$ из \eqref{eq3_3}. Получим критические значения \eqref{eq3_2},
занумерованные в порядке возрастания.
\begin{figure}[ht]
\centering
\includegraphics[width=70mm, keepaspectratio]{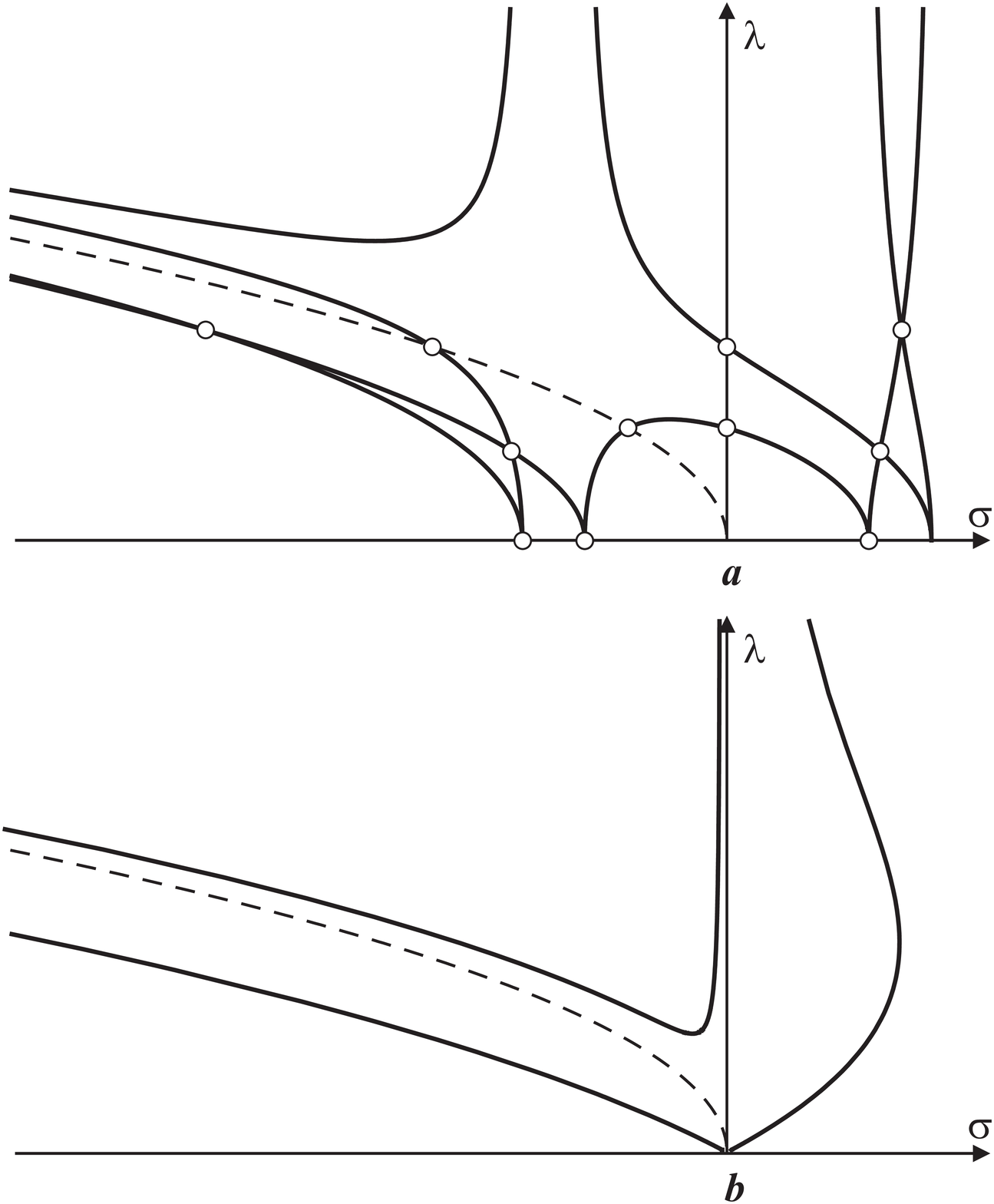}\\
\caption{Первое и второе разделяющие множества.}\label{fig1}
\end{figure}

Геометрия ПРМ определяется следующим непосредственно проверяемым
утверждением.
\begin{proposition}
Первое разделяющее множество имеет ровно четыре кратные точки при
$\lambda>0$
\begin{equation}\notag
    \begin{array}{c}
Q_1 \cap Q_2 = \{q_{+}^{(12)}, q_{-}^{(12)}\}, \qquad Q_1 \cap Q_3
=\{q_{+}^{(13)}, q_{-}^{(13)}\},
\end{array}
\end{equation}
где
\begin{equation}\label{eq3_5}
    \begin{array}{c}
q_{\pm}^{(12)} = (-a\pm \sqrt{4a^2+b^2},\sqrt{2a}), \qquad
q_{\pm}^{(13)} = (-b \pm \sqrt{a^2+4b^2},\sqrt{2b}),
    \end{array}
\end{equation}
четыре точки пересечения с осью $\lambda=0$ $($каждая принадлежит двум кривым$)$
\begin{equation}\notag
\begin{array}{c}
    Q_1 \cap Q_4 \cap \{\lambda=0\} = \{(-a+b,0),(a-b,0)\}, \\[2mm]
    Q_2 \cap Q_3 \cap \{\lambda=0\} = \{(-a-b,0),(a+b,0)\},
\end{array}
\end{equation}
две точки пересечения с осью $\sigma=0$
\begin{equation}\notag
\begin{array}{c}
    Q_3 \cap \{\sigma=0 \} = \left(0,\displaystyle{\frac{a+b}{\sqrt{a-b}}}\right), \qquad
        Q_4 \cap \{\sigma=0 \} = \left(0,\displaystyle{\frac{a-b}{\sqrt{a+b}}}\right)
\end{array}
\end{equation}
и две точки пересечения с особой параболой
\begin{equation} \notag
\begin{array}{lll}
Q_3 \cap \Pi = \displaystyle{\left(-\frac{(a+b)^2}{a-b},
\frac{a+b}{\sqrt{a-b}}\right),}\qquad  Q_4 \cap \Pi =
\displaystyle{\left(-\frac{(a-b)^2}{a+b},
\frac{a-b}{\sqrt{a+b}}\right).}
\end{array}
\end{equation}
\end{proposition}

Из найденных выражений для координат кратных точек следует, что
пересечения трех кривых $Q_i$ в одной точке невозможны. Таким
образом, множество $\mS_1$ вместе с отмеченным ранее особым
множеством $\mS_0$ разбивает верхнюю полуплоскость на 19 областей
(рис.~\ref{fig1},\,{\it a}, штриховой линией показана особая
парабола).

\section{Второе разделяющее множество}\label{sec4} Количество
решений в зависимости от параметров может изменяться при переходе
через такие значения, при которых уравнение \eqref{eq2_2} имеет
кратный корень по $u$. Назовем это множество параметров вторым
разделяющим множеством (ВРМ) и обозначим через $\mS_2$. Из условия
$L'_u=0$, учитывая, что в \eqref{eq2_2} $u\ne 0$, получим
\begin{equation}\label{eq4_1}
\begin{array}{l}
 5 u^3 \lambda^2 (\lambda^2 + \sigma)^2 - 4 u^2 \sigma (\lambda^2 + \sigma) [
\sigma (\lambda^2 + \sigma)^3-2 p^2 \lambda^4 ] +\\
 \qquad +3 r^4 u \lambda^6
\sigma^2 +4 r^4 \lambda^4 \sigma^4 (\lambda^2 + \sigma)^2 =0.
\end{array}
\end{equation}
Условие совместности по $u$ уравнений \eqref{eq2_2}, \eqref{eq4_1}
за пределами $\mS_0$ приводит к уравнению
\begin{equation} \label{eq4_2}
\begin{array}{l}
R_L=27 r^8 \lambda^{16} - 256 \sigma ( \lambda ^2 +\sigma)^3
[\sigma ( \lambda ^2 +\sigma )^3 - 2 p^2 \lambda ^4 ] ^3 + \\[1.5mm]
\qquad + 864 r^4 \lambda ^8 \sigma ( \lambda ^2 +\sigma ) ^3 [5
\sigma ( \lambda ^2 +\sigma ) ^3-2 p^2
 \lambda ^4]=0.
\end{array}
\end{equation}
Для его упрощения заметим, что в выражении $R_L$ параметры
$\lambda,\sigma$ входят в виде определенной комбинации. Обозначим
\begin{equation}\label{eq4_3}
\displaystyle{V=\frac{4(\lambda^2+\sigma)^3 \sigma}{\lambda^4},}
\end{equation}
тогда уравнение (\ref{eq4_2}) перепишется в виде
\begin{equation}\label{eq4_4}
R_L^*=V^4-24 p^2 V^3 +6(32p^4-45r^4)V^2-16p^2(32p^4-27r^4)V-27r^8=0.
\end{equation}

\begin{proposition}
В области параметров \eqref{eq2_1} уравнение \eqref{eq4_4} имеет
ровно два вещественных корня $V_1, V_2$ противоположных знаков
\begin{equation}\label{eq4_5}
    \begin{array}{c}
      V_1 = (\varkappa_1 +\varkappa_2)^3>0, \qquad V_2 = (\varkappa_1 - \varkappa_2)^3<0.
    \end{array}
\end{equation}
Здесь $\displaystyle{\varkappa_1=\sqrt{r^{4/3}+(p^4-r^4)^{1/3}\vphantom{\varkappa_1^{-1}}}}$, $\displaystyle{\varkappa_2=\sqrt{2r^{4/3}-(p^4-r^4)^{1/3}+ 2p^2 \varkappa_1^{-1} }}$.
\end{proposition}
\begin{proof} Непосредственное вычисление корней обозримых выражений
не дает. Выполним подстановку $V=W^3, r=q^3$, получим разложение ${R_L^* = P_1
P_2}$, где
\begin{equation}\notag
\begin{array}{l}
      P_1 = 3 q^8 + 8 p^2 W + 6 q^4 W^2 - W^4, \\[3mm]
      P_2= 9 q^{16} - 24 p^2 q^8 W +
   64 p^4 W^2 - 18 q^{12} W^2 - 48 p^2 q^4 W^3 + 39 q^8 W^4 -\\
   \qquad - 16 p^2 W^5 + 6 q^4 W^6 + W^8.
    \end{array}
\end{equation}
Дискриминант $P_2$ равен $r^{40/3} (512 p^4 - 169 r^4)^2 (p^4 -
r^4)^4$ (см. замечание~1), поэтому в области \eqref{eq2_1} многочлен $P_2$ кратных корней не имеет, а
численная проверка дает все комплексные корни. Вычисление корней
$P_1$ приводит к выражениям
\begin{equation}\notag
    \begin{array}{c}
      \displaystyle{
      \varepsilon_1 \sqrt{r^{4/3}+(p^4-r^4)^{1/3}}+\varepsilon_2\sqrt{2r^{4/3}-(p^4-r^4)^{1/3}+\varepsilon_1
      \frac{2p^2}{\sqrt{r^{4/3}+(p^4-r^4)^{1/3}}}}
      },
    \end{array}
\end{equation}
где $\varepsilon_i=\pm 1$. Заметим, что
$$
\left[\frac{2p^2}{\sqrt{r^{4/3}+(p^4-r^4)^{1/3}}}\right]^2-\left[2r^{4/3}-(p^4-r^4)^{1/3}\right]^2=3
(p^4 - r^4)^{2/3} > 0,
$$
поэтому вещественные корни отвечают только значению
$\varepsilon_1=1$. Произведение всех корней отрицательно, поэтому
пара вещественных корней имеет разные знаки, но, очевидно, что
корень с $\varepsilon_2=1$ положительный. Утверждение доказано.
\end{proof}

Теперь кривые в составе ВРМ легко строятся численно
(рис.~\ref{fig1},\,{\it b}). Удобно ввести параметризацию, положив
$\sigma= s \lambda^2$. Тогда из \eqref{eq4_3} получим следующие три
ветви ВРМ:
\begin{equation}\label{eq4_8}
\begin{array}{lll}
D_A:  \; \displaystyle{ \lambda = \sqrt[4]{\frac{V_1}{4s(1+s)^3}}},
& \displaystyle{ \sigma = \frac{s}{2}\sqrt{\frac{V_1}{s(1+s)^3}}},
&
s \in (-\infty,-1);\\
D_B: \; \displaystyle{ \lambda = \sqrt[4]{\frac{V_1}{4s(1+s)^3}}},
& \displaystyle{ \sigma = \frac{s}{2}\sqrt{\frac{V_1}{s(1+s)^3}}},
& s \in (0,+\infty); \\
D_C: \; \displaystyle{ \lambda = \sqrt[4]{\frac{V_2}{4s(1+s)^3}}},
& \displaystyle{ \sigma = \frac{s}{2}\sqrt{\frac{V_2}{s(1+s)^3}}},
& s \in (-1,0).
\end{array}
\end{equation}
Как видно из рис.~\ref{fig1},\,{\it b} кривые в составе $\mS_2$ вместе
с множеством $\mS_0$ делят верхнюю полуплоскость на шесть областей.
В каждой из них сохраняется количество корней многочлена $L(u)$.
Заметим, что $L(0)=-r^8\lambda^8 \sigma^6$ обращается в нуль только
на осях координат. Поэтому в каждой из областей не могут изменяться
и знаки корней. Теперь всю необходимую информацию о корнях $L(u)$
легко получить численно.

Для нахождения значения энергии, отвечающего этим разделяющим
случаям, необходимо выразить общий корень $u$ уравнений
\eqref{eq2_2}, \eqref{eq4_1}. Для этого в дополнение к \eqref{eq4_3}
выполним подстановку $u =\sigma \sqrt[3]{\lambda^2 \sigma} Z$.
Система уравнений примет вид
\begin{equation}\label{eq4_10a}
    \begin{array}{l}
      (2V)^{1/3} Z^4 \bigl[ 8p^2-V+2(2V)^{1/3}Z \bigr]+ 4r^4 \bigl[(2V)^{2/3}+2Z
      \bigr]Z^2-8r^8=0,\\[3mm]
      (2V)^{1/3}Z^2 \bigl[16 p^2- 2 V+5(2V)^{1/3}Z \bigr]+ 4 r^4 \bigl[(2V)^{2/3}+3Z\bigr]=0.
    \end{array}
\end{equation}
Условие ее совместности по $Z$, естественно, дает уравнение
\eqref{eq4_4} с решениями~\eqref{eq4_5}. Приравняем к нулю
результант левых частей \eqref{eq4_10a} по переменной $V^{1/3}$, получим
\begin{equation}\notag
    r^4(p^4-r^4)Z^{14}(Z^3-4r^4)^4=0,
\end{equation}
а поскольку из первого уравнения \eqref{eq4_10a} следует, что $Z\ne 0$, то $Z=r\sqrt[3]{4r}$, и,
следовательно, кратный корень $L(u)$ равен $ u = r \sigma \sqrt[3]{4
r \lambda^2 \sigma}$. Подставляя это значение вместе с \eqref{eq4_8}
в \eqref{eq3_4}, найдем
\begin{equation}\label{eq4_11}
    h=-\displaystyle{\frac{s(2s+1)V^{2/3}+(2s^2+s-1)r^{4/3}}{4\sqrt{s(1+s)^3}V^{1/6}}}.
\end{equation}
Здесь $V=V_{1,2}$ согласно \eqref{eq4_5}, а промежуток изменения параметра $s$ определяется
в соответствии с уравнениями \eqref{eq4_8}. Из \eqref{eq4_8}
возьмем формулы и для $\lambda(s)$. Получим, после перехода к
безразмерным переменным $b/a,h/a,\lambda/\sqrt{a}$, параметрические
уравнения разделяющей поверхности для классификации бифуркационных
диаграмм на изоэнергетических уровнях, разрешающие неявные уравнения
работы \cite{Ryab37}. Более того, промежутки изменения $s$
определяют условия, при которых действительно реализуются
перестройки таких диаграмм. В силу ограниченности объема публикации,
мы не приводим иллюстраций, которые теперь легко могут быть
построены.

\section{Пересечения разделяющих множеств}\label{sec5} Для того
чтобы получить общую картину на плоскости $(\sigma,\lambda)$, найдем
пересечения $\mS_2$ с кривыми $Q_i$, составляющими $\mS_1$.

\begin{figure}[ht]
\centering
\includegraphics[width=90mm,keepaspectratio]{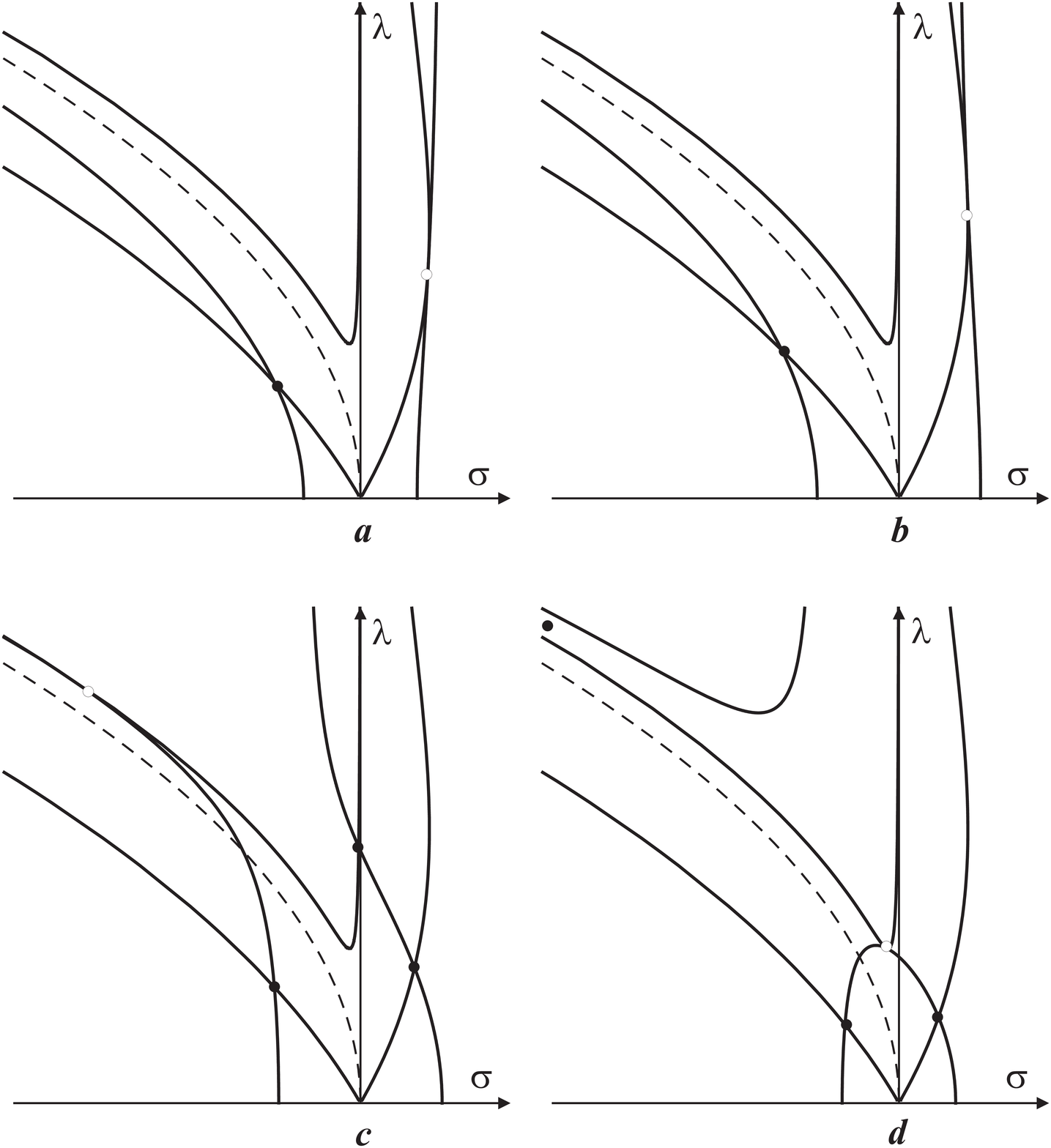}
\caption{Пересечения $\mS_2 \cap Q_i$.}\label{fig2}
\end{figure}

Остановимся более подробно на пересечении $\mS_2 \cap Q_1$. Из
уравнения $Q_1$ выразим
\begin{equation} \label{eq5_1}
\displaystyle{\lambda^2 = \frac {\sigma^2 - (a - b)^2}{a + b -
\sigma}}
\end{equation}
и подставим в уравнение (\ref{eq4_2}). Получим $M_1^2(\sigma)
N_1(\sigma) = 0$, где
\begin{equation} \notag
\begin{array}{l}
M_1 = (a - b) ^4 (a + b) - 4 (a ^2 - b ^2) ^2 \sigma +  6 (a - b) ^2
(a + b) \sigma ^2 -  \\
\phantom{M_1=} -4 (a - b) ^2 \sigma ^3 +(a + b) \sigma ^4, \\
N_1 = (a-b)^{10}\sum\limits_{i=0}^{12} (a-b)^{-i} A_i \sigma^i
\end{array}
\end{equation}
и
\begin{equation}\notag
\begin{array}{l}
A_0 = 27 (a - b) ^{2} (a + b)^6, \\[1.5mm]
A_1=-4 (a + b) (53 a ^6 - 270 a ^5 b + 1131 a ^4 b ^2 -1188 a ^3 b
^3 + 1131 a ^2 b ^4 -\\[1.5mm]
\phantom{M_1 =}- 270 a b ^5 + 53 b
^6),\\[1.5mm]
A_2= 2 (a + b) ^2 (331 a ^6 - 3342 a ^5 b + 9525 a ^4 b ^2 - 12516 a
^3 b ^3 + 9525 a ^2 b ^4 -\\[1.5mm]
\phantom{M_1 =}-
 3342 a b ^5 + 331 b ^6), \\[1.5mm]
A_3= -12 (a^2 - b^2) (75 a ^6 - 286 a ^5 b -
 1547 a ^4 b ^2 + 2236 a ^3 b ^3 -\\[1.5mm]
\phantom{M_1 =} - 1547 a ^2 b ^4 - 286 a b ^5 +
 75 b ^6), \\[1.5mm]
A_4= - (75 a ^8 -
 14492 a ^7 b + 39404 a ^6 b ^2 + 12444 a ^5 b ^3 -50286 a ^4 b ^4 + \\[1.5mm]
\phantom{M_1 =}+12444 a ^3 b ^5 + 39404 a ^2 b
 ^6 - 14492 a b ^7 + 75 b ^8), \\[1.5mm]
A_5=  8 (a^2 - b^2) (267 a ^6 - 2986 a ^5 b -
 1627 a ^4 b ^2 + 6068 a ^3 b ^3  -1627 a ^2 b ^4 -\\[1.5mm]
 \phantom{M_1 =} - 2986 a b ^5 + 267 b ^6),  \\[1.5mm]
A_6 = - 12 (273 a ^8 - 5492 a ^6 b ^2 +
 6342 a ^4 b ^4 - 5492 a ^2 b ^6 +
 273 b ^8),\\[1.5mm]
A_{12-i}(a,b)=A_i(a,-b) \qquad (i=0,\ldots,5).
\end{array}
\end{equation}
Корни $M_1$ отвечают за точки касания кривых $\mS_2$ и $Q_1$, а
корни $N_1$ отвечают за точки трансверсального пересечения кривых
$\mS_2$ и $Q_1$. Из этих корней необходимо взять лишь те, которые
обеспечивают вещественный корень уравнения \eqref{eq5_1}
относительно $\lambda$, т.е. удовлетворяющие неравенству
\begin{equation}\label{eq5_3}
\begin{array}{l}
P_1(\sigma)=[\sigma^2 - (a - b)^2](a + b - \sigma) \geqslant 0.
\end{array}
\end{equation}
Количество и взаимное расположение таких корней может измениться
лишь в следующих случаях: один из многочленов $M_1,N_1$ имеет общий
корень с $P_1$, либо один из этих многочленов имеет кратный корень в
области \eqref{eq5_3}. Вычисляем результанты
\begin{equation}\notag
    \begin{array}{l}
      \Res(M_1,P_1,\sigma) = -a^4 b ^4 (a-b)^6 (a+b);\\
      \Res(N_1,P_1,\sigma) = -a^{16} b ^{16} (a-b)^{20} (a+b)^2; \\
      \Res(M_1,M_1',\sigma) = -a^4 b ^4 (a-b)^8 (a+b)^3; \\
      \Res(N_1,N_1',\sigma) = a^{66} b^{66}(a-b)^{106}(a+b)^{16} (343 a^4+ 1362 a^2
b^2+343 b^4)^4.
    \end{array}
\end{equation}
Самым удивительным является, конечно, последний. В нем отброшенный
здесь числовой множитель содержит 76 цифр и равен
$2^{212}{\times}3^{24}$, однако, вычисляется точно. Ни одно из
полученных выражений в нуль не обращается, поэтому качественная
структура множества $\mS_2 \cap Q_1$  не зависит от параметров $a,b$
в выбранной области их изменения. Следовательно, численное решение
полностью определяет и качественную картину. Находим, что $M_1$
имеет единственный корень в области \eqref{eq5_3} и он дает точку
касания $Q_1$ с кривой $D_B$. Многочлен $N_1$ также имеет
единственный корень в указанной области. Он порождает точку
трансверсального пересечения $Q_1$ с кривой $D_A$
(рис.~\ref{fig2},\,{\it a}).

Рассмотрим пересечение $\mS_2 \cap Q_2$. Отметим, что все уравнения
получаются из уравнений для $Q_1$ заменой $b  \to -b $, поэтому
верны выводы об отсутствии каких-либо общих или кратных корней у
соответствующих многочленов $P_2,M_2,N_2$, т.е. качественная
картина от параметров снова не зависит. Находим, что кривая $Q_2$
имеет ровно одну точку пересечения с $\mS_2$ на ветви $D_A$ и ровно
одну точку касания с ветвью $D_B$ (рис.~\ref{fig2},\,{\it b}).

Замечая, что случай с кривой $Q_3$ получается из предыдущего заменой
${\lambda^2 \to -\lambda^2}$, ${\sigma \to -\sigma}$, с помощью
аналогичных выкладок и рассуждений находим, что кривая $Q_3$ имеет
ровно по одной точке пересечения с каждой из ветвей $D_A,D_B,D_C$ и
одну точку касания с ветвью $D_C$ (рис.~\ref{fig2},\,{\it c}).

Для пересечения $\mS_2 \cap Q_4$ все уравнения получаются из
уравнений для $Q_3$ заменой ${b  \to -b}$. Находим, что кривая $Q_4$
имеет ровно по одной точке пересечения с каждой из ветвей
$D_A,D_B,D_C$ и одну точку касания с ветвью $D_C$. Точка $Q_4 \cap
D_C$ имеет отрицательную координату $\sigma$ по модулю на порядки
больше остальных (например, для $a=1$, $b=0.5$ эта точка имеет
координаты $\sigma \approx -1104.12; \; \lambda \approx 33.25$),
поэтому на рис.~\ref{fig2},\,{\it d} она показана условно.

\section{Перестройки разделяющего множества}\label{sec6} Для почти
всех значений физических параметров $a,b$ структура разделяющего
множества $\mS$ полностью определена.
Перестройки этой структуры по указанным параметрам могут отвечать
лишь случаям наличия в составе разделяющего множества точек
кратности три и выше. Как отмечено ранее, пересечения трех $Q_i$
всегда пусты, поэтому точек кратности четыре не существует, а точки
кратности три возможны лишь для таких значений параметров $a,b$, при
которых для некоторой пары индексов $i \neq j$
\begin{equation} \label{eq6_1}
\mS_2 \cap Q_i \cap Q_j \neq \varnothing.
\end{equation}
Заметим, что это же условие является необходимым и достаточным для
того, чтобы при переходе параметров $a,b$ через заданные значения
могло измениться и взаимное расположение точек двух различных
множеств $\mS_2 \cap Q_i$.

\begin{figure}[ht]
\centering
\includegraphics[width=90mm, keepaspectratio]{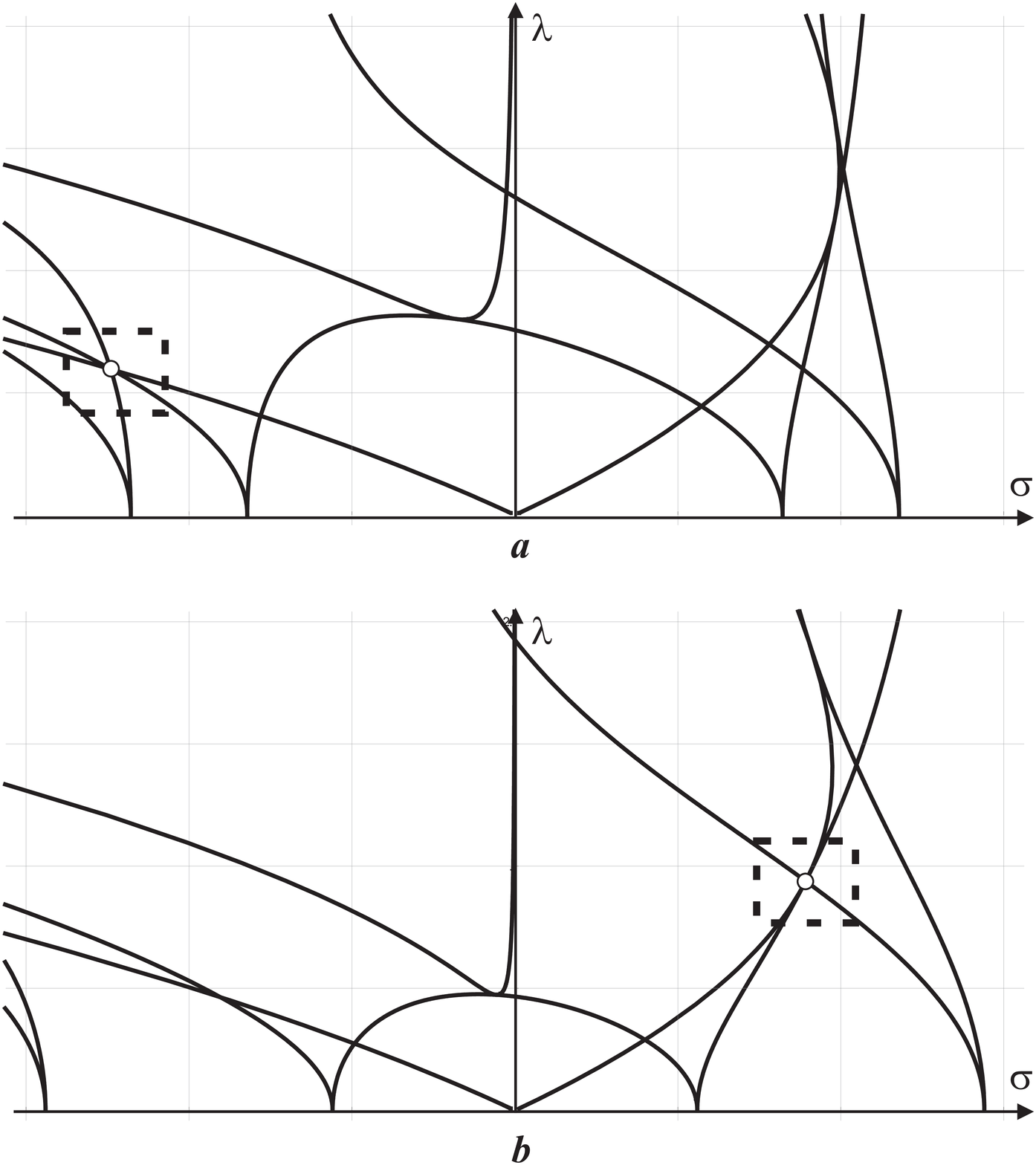}
\caption{Перестройки областей в окрестности
$b/a=\gamma_*,\gamma^*.$}\label{fig3}
\end{figure}

Согласно предложению 1 имеется всего четыре точки, принадлежащие при
${\lambda > 0}$ паре множеств $Q_i$, а именно, точки \eqref{eq3_5}.
Для исследования случая (\ref{eq6_1}) необходимо проверить
возможность $q_{\pm}^{(ij)} \in \mS_2$, что означает равенство
$R_L(q_{\pm}^{(ij)})=0$. Имеем
\begin{equation}\notag
R_L(q_{\pm}^{(13)}) = 1024\, a^{20}\, (A_1 \pm A_2)\,X, \qquad
X\,=(b/a)^2>0,
\end{equation}
где
\begin{equation}\notag
\begin{array}{l}
A_1 = 1+32 X +118 X^2 - 1756 X^3- 10511 X^4 +8388 X^5
+\\[1.5mm]
\phantom{A_1 = }+143104 X^6+221840 X^7-768 X^8);\\[1.5mm]
A_2=8 (1 +3 X) \sqrt{\mathstrut 1+4 X}(1+10 X-60X^2-370 X^3+891 X^4+\\[1.5mm]
\phantom{A_1 = }+4632 X^5+16 X^6).
\end{array}
\end{equation}
Обозначая $\gamma=b/a \in [0,1]$, вычислим
\begin{equation}\label{eq6_2}
\begin{array}{l}
A_1^2-A_2^2=-(1+4 \gamma+9\gamma^2+18\gamma^3+20\gamma^4+12 \gamma^5)^2 \times \\[1.5mm]
\phantom{A_1^2-A_2^2=}\times (-1+4 \gamma-9\gamma^2+18\gamma^3-20\gamma^4+12 \gamma^5)^2\times \\[1.5mm]
\phantom{A_1^2-A_2^2=}\times (-1+4 \gamma^2+602 \gamma^4+7140
\gamma^6+24863 \gamma^8+32928 \gamma^{10}).
\end{array}
\end{equation}
Уравнение $ A_1^2-A_2^2=0$ уже не содержит параметров и имеет ровно
два положительных корня
\begin{equation}\label{eq6_3}
\begin{array}{l}
b/a = \gamma_* \approx 0.17865486876 \quad
({\textrm{точка}}\;q_-^{(13)}, \; \sigma < 0), \\
b/a = \gamma^* \approx 0.44041017065 \quad
({\textrm{точка}}\;q_+^{(13)}, \; \sigma>0).
\end{array}
\end{equation}
При этом первый из них является корнем многочлена в последней
скобке \eqref{eq6_2}, т.е. простым корнем, и поэтому в соответствующем случае
все три кривые пересекаются трансверсально, а второй является корнем
многочлена во второй скобке (\ref{eq6_2}), входящей в это уравнение
в квадрате, и поэтому, являясь кратным корнем, связан с точками
касания.  Разделяющее множество для значений (\ref{eq6_3}) показано
на рис.~\ref{fig3}. Видно, что первое значение (рис.~\ref{fig3},\,{\it
a}) отвечает случаю, когда кривые $Q_1$ и $Q_3$ пересекаются на
ветви $D_A$, второе (рис.~\ref{fig3},\,{\it b}) -- случаю, когда точка
касания кривых $D_C$ и $Q_1$ находится на кривой $Q_3$, что и
объясняет кратность данного значения $b/a$ в
уравнении~(\ref{eq6_2}).

Заметим теперь, что точки $q_{\pm}^{(12)}$ получаются из
$q_{\pm}^{(13)}$ перестановкой значений $a$ и~$b$, а функция $R_L$
от такой перестановки не изменяется. Поэтому, проводя аналогичные
вычисления для точек $q_{\pm}^{(12)}$, придем к уравнению вида
\eqref{eq6_2}, в котором уже $Y=a/b$. Для этого отношения найденные
выше корни являются посторонними, так как $a/b > 1$. Таким образом,
этот случай не реализуется при допустимых значений параметров.

Итак, построенное множество $\mS$ разбивает
верхнюю полуплоскость параметров $(\sigma,\lambda)$ на 35 открытых
областей, которые несложно закодировать в соответствии с номерами
пары областей, порожденных первым и вторым разделяющими множествами.
При переходе через значения \eqref{eq6_3} одна из таких областей
исчезает, взамен появляется область уже с другим кодом и,
соответственно, с другим количеством порожденных периодических
траекторий. Поскольку внутри каждой из образовавшихся областей
количество траекторий измениться уже не может, то численный анализ
дает полную картину. В четырех областях движений нет, в остальных
количество траекторий варьируется от одной до пяти. Исследование
разделяющего множества, определяющего условия существования и
количества периодических траекторий закончено.

\section{Заключение} В работе представлено полное аналитическое
исследование разделяющих множеств в пространстве параметров задачи о
движении гиростата Ковалевской в двойном поле, связанных с наличием
и бифуркациями внутри совокупности критических точек ранга 1
интегрального отображения. Образ этого множества в трехмерном
пространстве существенных параметров изоэнергетических многообразий
определяет атлас бифуркационных диаграмм. Получены явные уравнения
\eqref{eq3_2} и \eqref{eq4_11} для поверхностей в составе такого
атласа, фактически решающие неявные уравнения высокой степени,
полученные в работе \cite{Ryab37} и, что еще более важно,
исключающие посторонние решения этих уравнений. В результате
получена аналитическая основа для полной классификации и
визуализации бифуркационных диаграмм гиростата Ковалевской в двойном
поле.

%\clearpage

\end{document}